\documentclass[aps,twocolumn,showpacs,10pt]{revtex4-1}
\usepackage{tensor,amsmath}

\begin{document}
\title{Resolution of the Mansuripur Paradox}
\author{Daniel J. Cross}
\affiliation{Physics Department, Bryn Mawr College, Bryn Mawr, Pennsylvania, 19010, USA}
\date{\today: To be submitted, PRL}
\begin{abstract}
The interaction of a magnetic dipole with a point charge leads to an apparent paradox when analyzed using the 3-vector formulation of the Lorentz force.  Specifically, the dipole is subject to a torque in some frames and not in others.  We show that when analyzed according to the covariant 4-vector formulation the paradox disappears.  The torque that arises in certain frames is connected to the time-space components of the torque in the rest frame, giving rise to ``hidden'' momentum.
\end{abstract}
\maketitle
\paragraph*{Introduction.}%
In a recent Letter~\cite{Man2012} Mansuripur presents a paradox where the Lorentz force law of electrodynamics fails to accord with the principle of relativity.  The paradox is reminiscent of the Trouton-Noble~\cite{TN,vL} and right-angle lever paradoxes~\cite{NM} (see also ~\cite{J2004}).  Mansuripur considers a point charge $q$ at the origin and an electrically neutral point magnetic dipole ${\bf m}=m\hat{\bf x}$ at ${\bf r}=d\hat{\bf z}$ in their common rest frame $S$.  In this frame the dipole experiences neither a force nor a torque from the point charge.  However, when viewed from a frame $S'$ moving in the negative $z$-direction, the magnetic dipole gains an electric dipole moment, which the Lorentz law predicts will experience a torque from the electric field of the point charge.  Mansuripur then argues that the Lorentz force law should be abandoned in favor of, e.g.\ the Einstein-Laub law, which predicts no torque in all frames.  In this Letter we argue that the torque predicted in the moving frame is correct and necessary to balance the ``hidden'' angular momentum of the moving dipole~\cite{S}.  The existence of this hidden angular momentum comes directly from the covariant formulation of the Lorentz law.  In fact, it arises from a hidden linear momentum in the rest frame that is generated by the time-space components of the torque tensor.  In the 3-vector formulation it is this connection that is hidden.

\paragraph*{Non-Covariant Analysis and the Paradox.}
We begin by reviewing the non-covariant description according to both the Lorentz and Einstein-Laub prescriptions.  Consider a neutral point particle with magnetic dipole moment ${\bf m}$ and electric dipole momentum ${\bf p}$ moving with coordinate velocity ${\bf v}$.  The moment densities are ${\bf M}={\bf m}\delta({\bf r}-{\bf r}_0-{\bf v}t)$ and ${\bf P}={\bf p}\delta({\bf r}-{\bf r}_0-{\bf v}t)$, where ${\bf r}_0$ is the displacement of the dipole from the origin at time zero.  In the usual formulation these moment densities become bound charge and current densities $\rho=-\nabla\cdot{\bf P}$ and ${\bf J}=\partial{\bf P}/\partial t+\nabla\times{\bf M}$.  The Lorentz force is obtained by integrating the force density ${\bf f}_L=\rho{\bf E}+{\bf J}\times{\bf B}$, yielding (using the properties of delta functions)
$${\bf F}_L=({\bf p}\cdot\nabla){\bf E}+\nabla({\bf m}\cdot{\bf B})+
({\bf v}\cdot\nabla)({\bf p}\times{\bf B}),$$
where the subscript $L$ stands for Lorentz.  The expression for the Einstein-Laub force density is (with no free charge/current)
$${\bf f}_E=({\bf P}\cdot\nabla){\bf E}+\partial{\bf P}/\partial t\times {\bf B}
+({\bf M}\cdot\nabla){\bf B}-\partial{\bf M}/\partial t\times {\bf E},
$$
which integrates to
$${\bf F}_E=({\bf p}\cdot\nabla){\bf E}+({\bf m}\cdot\nabla){\bf B}+
({\bf v}\cdot\nabla)({\bf p}\times{\bf B}-{\bf m}\times{\bf E}),$$
where the subscript $E$ stands for Einstein-Laub.  These two force expressions have two terms in common.  However, since ${\bf m}$ is constant, the remaining term in the Lorentz force can be written 
$$\nabla({\bf m}\cdot{\bf B})=({\bf m}\cdot\nabla){\bf B}+
{\bf m}\times(\nabla\times{\bf B}).$$
Finally, using Amp\`ere's Law and the fact that in the moving frame $\partial{\bf E}/\partial t=-{\bf v}\cdot\nabla{\bf E}$, 
the two integrated force expressions are exactly equal here.  Not only do both force laws both give zero, they give equivalent expressions.

The Lorentz torque is found by integrating the torque density ${\bf r}\times{\bf f}_L$, yielding
$${\bf T}_L={\bf r}\times{\bf F}_L+{\bf p}\times{\bf E}+{\bf m}\times{\bf B}
+{\bf v}\times({\bf p}\times{\bf B}).$$
When the dipole is viewed from the moving frame every term vanishes except for ${\bf p}'\times{\bf E}'$, which gives the offending torque.  Using the standard transformations~\cite{J} we find  ${\bf p}'=\gamma{\bf v}\times{\bf m}=\gamma vm\hat{\bf y}'$ and ${\bf E}'={\bf E}$ (at the dipole), so the torque is $\gamma vm E\hat{\bf x}'$.

The torque density for Einstein-Laub is supplemented by additional terms and reads ${\bf r}\times{\bf f}_E+{\bf P}\times{\bf E}+{\bf M}\times{\bf B}.$
It integrates to
$${\bf T}_E={\bf r}\times{\bf F}_E+
{\bf p}\times{\bf E}+{\bf m}\times{\bf B}
+{\bf v}\times({\bf p}\times{\bf B}-{\bf m}\times{\bf E}).$$
In the moving frame the extra term $-{\bf v}\times({\bf m}'\times{\bf E}')$ exactly cancels the non-vanishing ${\bf p}'\times{\bf E}'$ term, giving zero torque for the moving dipole.

\paragraph*{Covariant Analysis and the Resolution.}
How can a covariant force law give rise to a non-trivial torque after a Lorentz transformation?  Here we analyze the situation in detail.  We set $\epsilon_0=\mu_0=c=1$, and our metric has signature $(+,-,-,-)$.  The Lorentz force law can be written in covariant form, even for bound charges and currents~\cite{J}.  First, the electric and magnetic moment densities form the second rank tensor
$$\tensor{M}{^\alpha^\beta}=
\begin{pmatrix}
0 & P_x & P_y & P_z\\
-P_x & 0 & -M_z & M_y\\
-P_y & M_z & 0 & -M_x\\
-P_z & -M_y & M_x & 0
\end{pmatrix}.$$
For short we write $M=({\bf P},-{\bf M})$ in terms of the defining 3-vectors.
With the bound current density 4-vector defined as the divergence $j^\beta=\partial_\alpha\tensor{M}{^\alpha^\beta}$, the force law is
$$f^\alpha=\tensor{F}{^\alpha_\beta}j^\beta,$$
where $F=(-{\bf E},-{\bf B})$ is the Faraday tensor in the same notation as above.  For a point dipole $\tensor{M}{^\alpha^\beta}=\tensor{m}{^\alpha^\beta}\delta({\bf r}-{\bf r}_0-{\bf v}t)$, where $m=({\bf p},-{\bf m})$ is the tensor of moments.

The spatial components of $f^\alpha$ give the force 3-vector as expected.  The time component works out to
$$f^0={\bf E}\cdot\left(\frac{\partial{\bf P}}{\partial t}+\nabla\times{\bf M}\right),$$
which integrates to
$$F^0={\bf v}\cdot\nabla({\bf P}\cdot{\bf E})+{\bf m}\cdot(\nabla\times{\bf E}).$$
The entire four force vanishes for the dipole in both frames.

The torque density is not a vector, but the antisymmetric tensor
$$\tensor{t}{^\alpha^\beta}=x^\alpha f^\beta-x^\beta f^\alpha.$$
The space-space components are the usual torque components given previously.  However, this tensor has non-trivial time-space components.  If we write $T=({\bf R},{\bf T})$ for the integrated torque, then
$${\bf R}={\bf m}\times{\bf E}-{\bf p}\times{\bf B}
+t{\bf F}-{\bf r}F^0.$$
These ``torque'' components are connected with the motion of the center of energy~\cite{LL}.  Even in the rest frame this expression does not vanish for the dipole: ${\bf R}={\bf m}\times {\bf E}=-mE\hat {\bf y}$.  As we will see shortly, this corresponds to a constant ``hidden'' momentum in the negative $y$-direction due to the interaction of the magnetic moment with the electric field.  Under a Lorentz boost to the moving frame the space-space and time-space components mix.  In particular, the $x$-component of the torque after the boost is
$$T^{x'}\equiv
\tensor{T}{^'^y^z}=
\tensor{\Lambda}{^y_\alpha}
\tensor{\Lambda}{^z_\beta}
\tensor{T}{^\alpha^\beta}
=\tensor{\Lambda}{^z_t}
\tensor{T}{^y^t}
=-\gamma v{\bf R}^{y},
$$
where $\Lambda$ is the boost matrix.  For the dipole this yields
$$T^{'x}=\gamma vmE,$$
which is exactly the torque calculated previously.  Thus we see that the torque in the moving frame is connected to the hidden momentum in the rest frame.

To complete the discussion we need to consider the the angular momentum, which is the second rank tensor $\tensor{J}{^\alpha^\beta}=x^\alpha p^\beta-x^\beta p^\alpha,$ with $p$ the 4-momentum.  The equation of motion is
$$\frac{d}{d\tau}\tensor{J}{^\alpha^\beta}=\tensor{T}{^\alpha^\beta}.$$
If we write $J=({\bf N},{\bf L})$ we have the 3-vector ${\bf N}=t{\bf p}-E{\bf r}$, where $E=p^0$ is the total energy.  But in the rest frame $F^0$ vanishes so no work is done and $E$ is constant.  But ${\bf r}$ is also constant in the rest frame and the equation of motion becomes
${\bf R}=d(t{\bf p})/dt$ or ${\bf m}\times{\bf E}={\bf p},$ giving the hidden momentum as claimed.  This momentum can also be obtained by integrating the stress-energy tensor of a current loop representing the dipole~\cite{V}.  A non-zero value is obtained because the interaction with the electric field distorts the stress-energy distribution.

Finally, upon transforming to the moving frame this constant hidden momentum becomes a non-constant hidden angular momentum ${\bf J}'=\gamma v mE \tau\hat {\bf x}'$  ($t=\tau$).  There must then be a torque $d{\bf J}/d\tau=\gamma v mE\hat {\bf x}'$.  This is, of course, precisely what is obtained by transforming the space-time torque from the rest frame.  Thus we conclude that while the torque in the moving frame exists, it merely balances the changing hidden angular momentum rather than causing a precession of the spin.  In this way the paradox is resolved.


\begin{thebibliography}{9}
\expandafter\ifx\csname natexlab\endcsname\relax\def\natexlab#1{#1}\fi
\expandafter\ifx\csname bibnamefont\endcsname\relax
  \def\bibnamefont#1{#1}\fi
\expandafter\ifx\csname bibfnamefont\endcsname\relax
  \def\bibfnamefont#1{#1}\fi
\expandafter\ifx\csname citenamefont\endcsname\relax
  \def\citenamefont#1{#1}\fi
\expandafter\ifx\csname url\endcsname\relax
  \def\url#1{\texttt{#1}}\fi
\expandafter\ifx\csname urlprefix\endcsname\relax\def\urlprefix{URL }\fi
\providecommand{\bibinfo}[2]{#2}
\providecommand{\eprint}[2][]{\url{#2}}

\bibitem[{\citenamefont{Mansuripur}(2012)}]{Man2012}
\bibinfo{author}{\bibfnamefont{M.}~\bibnamefont{Mansuripur}},
  \bibinfo{journal}{Phys. Rev. Lett.} \textbf{\bibinfo{volume}{108}},
  \bibinfo{pages}{193901} (\bibinfo{year}{2012}), \eprint{arXiv: 1205.0096}.

\bibitem[{\citenamefont{Trouton and Noble}(1903)}]{TN}
\bibinfo{author}{\bibfnamefont{F.~T.} \bibnamefont{Trouton}} \bibnamefont{and}
  \bibinfo{author}{\bibfnamefont{H.~R.} \bibnamefont{Noble}},
  \bibinfo{journal}{Proc. Royal Soc.} \textbf{\bibinfo{volume}{74}},
  \bibinfo{pages}{132} (\bibinfo{year}{1903}).

\bibitem[{\citenamefont{\lowercase{v}on Laue}(1912)}]{vL}
\bibinfo{author}{\bibfnamefont{M.}~\bibnamefont{\lowercase{v}on Laue}},
  \bibinfo{journal}{Annalen der Physik} \textbf{\bibinfo{volume}{343}},
  \bibinfo{pages}{370} (\bibinfo{year}{1912}).

\bibitem[{\citenamefont{Nickerson and McAdory}(1975)}]{NM}
\bibinfo{author}{\bibfnamefont{J.~C.} \bibnamefont{Nickerson}}
  \bibnamefont{and} \bibinfo{author}{\bibfnamefont{R.~T.}
  \bibnamefont{McAdory}}, \bibinfo{journal}{Am. J. Phys.}
  \textbf{\bibinfo{volume}{43}}, \bibinfo{pages}{615} (\bibinfo{year}{1975}).

\bibitem[{\citenamefont{Jackson}(2004)}]{J2004}
\bibinfo{author}{\bibfnamefont{J.~D.} \bibnamefont{Jackson}},
  \bibinfo{journal}{Am. J. Phys} \textbf{\bibinfo{volume}{72}},
  \bibinfo{pages}{1484} (\bibinfo{year}{2004}).

\bibitem[{\citenamefont{Shockley}(1968)}]{S}
\bibinfo{author}{\bibfnamefont{W.}~\bibnamefont{Shockley}},
  \bibinfo{journal}{Phys. Rev. Lett.} \textbf{\bibinfo{volume}{20}},
  \bibinfo{pages}{343} (\bibinfo{year}{1968}).

\bibitem[{\citenamefont{Jackson}(1999)}]{J}
\bibinfo{author}{\bibfnamefont{J.~D.} \bibnamefont{Jackson}},
  \emph{\bibinfo{title}{Classical Electrodynamics}}
  (\bibinfo{publisher}{Wiley}, \bibinfo{address}{New York, NY},
  \bibinfo{year}{1999}), \bibinfo{edition}{3rd} ed.

\bibitem[{\citenamefont{Landau and Lifshitz}(1975)}]{LL}
\bibinfo{author}{\bibfnamefont{L.}~\bibnamefont{Landau}} \bibnamefont{and}
  \bibinfo{author}{\bibfnamefont{E.~M.} \bibnamefont{Lifshitz}},
  \emph{\bibinfo{title}{The Classical Theory of Fields}}
  (\bibinfo{publisher}{Butterworth-Heinemann}, \bibinfo{address}{Burlington,
  MA}, \bibinfo{year}{1975}), \bibinfo{edition}{4th} ed.

\bibitem[{\citenamefont{Vanzella}(2012)}]{V}
\bibinfo{author}{\bibfnamefont{D.~A.~T.} \bibnamefont{Vanzella}}
  (\bibinfo{year}{2012}), \eprint{arXiv: 1205.1502}.

\end{thebibliography}

\end{document}